\documentclass{article}

\begin{document}

\def\square{\,\hbox{\vrule\vbox{\hrule\phantom{N}\hrule}\vrule}\,}

\thispagestyle{plain}
\def\upcirc#1{\vbox{\ialign{##cr
$\circ\!$cr \noalign{\kern-0.1 pt\nointerlineskip}
$\hfil\displaystyle{#1}\hfil$cr}}}
\def\i{\alpha}
\def\j{\beta}
\def\k{\gamma}
\def\m{\delta}

\title{Studying conformally flat spacetimes with an elastic stress energy tensor using 1+3 formalism}

\author {I. Brito
\\ Centro de Matem\' atica,
 \\ Universidade do Minho,
\\Portugal
\\
ireneb@math.uminho.pt
\and M.P. Machado Ramos
\\ Centro de Matem\' atica,
\\ Universidade do Minho,
 \\Portugal
\\
mpr@math.uminho.pt}

\maketitle

\section*{Abstract.}

Conformally flat spacetimes with an elastic stress energy tensor having diagonal trace-free anisotropic pressure are
investigated using 1+3 formalism. The 1+3 Bianchi and Jacobi identities and Einstein field equations are written for a particular case with
a conformal factor dependent on only one spatial coordinate. Solutions with non null anisotropic pressure are obtained.

\section{Introduction}

The theory of elasticity in the context of general relativity was developed in the mid twentieth century. The need for such a theory
 came in the late 1950s with Weber´s bar antenna for gravitational waves \cite{weber}, in order to explain how these waves interact with elastic solids.
 Actually, for this phenomena the weak-field approximation was sufficient in the treatment of the problem given by Weber. Only in 1973, in a paper
 by Carter and Quintana \cite{carter}, did a fully developed nonlinear theory of elasticity adapted to general relativity appear,
 remaining to this day a standard reference in the field, although the basic theoretical framework of this theory had already been given
 by Souriau in \cite{souriau}. Also before the article by Carter and Quintana, work by Maugin made considerable contributions to
 the field \cite{maugin1}-\cite{maugin2}. Lately, the theory of elasticity in general relativity was reconsidered
 by Magli and Kijowski \cite{magli1}-\cite{magli2} and Christodoulou \cite{christo}, in this work they explore the gauge character of relasticity.
 Authors such as Beig and Schmidt, have proven several existence and uniqueness theorems \cite{beig}. More recently, Karlovini and Samuelsson have
 given a self contained formulation of general relativistic elasticity in \cite{karl1}. They applied the theory of elasticity to spherically
 symmetric space-times and studied radial and axial perturbations \cite{karl2}-\cite{karl4}. Park \cite{park} established existence theorems
 for spherically symmetric static solutions for elastic bodies and Brito, Carot and Vaz \cite{irene} have obtained static shear-free and
 non-static shear-free solutions for spherically symmetric elastic spacetimes. Calogero and Heinzle \cite{calo} studied the dynamics of Bianchi type I
 elastic spacetimes.

In this work we study a very simple elastic model with a diagonal trace-free anisotropy pressure tensor using the 1+3 extended frame approach and
following the notation given in Uggla \cite{uggla}. We note that work on the covariant 1+3 splitting of fluid spacetime geometries was first
initiated by Eisenhart and Synge and continued by G$\ddot{o}$del, Raychaudhuri and other authors such as Sch$\ddot{u}$cking, Ehlers, Sachs and
Tr$\ddot{u}$mper (related references \cite{ehlers}, \cite{wain}, \cite{misner}, \cite{ellis}). In the paper by Uggla the basic dynamical equations of
the extended 1+3 orthonormal frame approach are explicitly given in terms of variables that are naturally adapted to the 1+3 structure, and they
include the Bianchi and Jacobi identities, the Einstein field equations and the commutators. This formulation is analogous to the Newmann Penrose
approach \cite{np} in the sense that a null congruence is replaced by a timelike congruence. The general properties of the 1+3 orthonormal frame
can be found in books such as Wald \cite{wald} and Felice and Clarke \cite{felice} and in Edgar \cite{edgar}. Several applications have been
discussed in Pirani \cite{pirani}, Ellis \cite{ellis1} and MacCallum \cite{mac}. A more complete list of references can be found in \cite{uggla}.

The organization of this paper is as follows. In section \ref{sec:formalism} we outline the theory of the 1+3 formalism and present the 1+3 split
of the commutators, curvature variables and their field equations, namely Bianchi and Jacobi identities and Einstein filed equations.
 We will follow the same notation convention for tensor indices as used in \cite{uggla}. Spacetime coordinate tensor indices will be denoted by
 letters from the second half of the Greek alphabet $(\mu , \nu , \rho, ...=0-3)$ while spatial coordinate indices are represented by letters from
 the second half of the Latin alphabet $(i,j,k,...=1,...,3)$. Orthonormal frame indices will be denoted by letters from the first half of
 the Latin alphabet $(a,b,c,...=0,...,3)$ while spatial frame indices are chosen from the first half of the Greek
 alphabet $(\alpha , \beta , \gamma, ...=1-3)$. In section \ref{sec:relasticity} we give a summary of the theory of relativistic elasticity.
 In section \ref{sec:conflat} we study conformally flat spacetimes with a conformal factor that depends only on one spatial coordinate with an elastic source given by a diagonal trace-free anisotropic pressure tensor with only one independent component,
 using the 1+3 formalism. We calculate the 1+3 equations for a very simple case and determine solutions with non null
 anisotropic pressure.

\section{1+3 Formalism}
\label{sec:formalism}
When studying a dynamical model in general relativity possessing an energy-momentum-stress tensor with a timelike eigendirection, for example perfect fluids, the associated timelike vector field $u$ on the spacetime $(M,g)$ determines the projection tensors $U$ and $h$, which project parallel and orthogonal to $u$ in the tangent space at each point $p\in (M,g)$, respectively. $u$ is chosen to be a unit timelike vector:

\begin{equation}
u_\mu u^\mu=-1,
\label{unitvector}
\end{equation}

\noindent and the projection tensors $U$ and $h$ are defined by
\begin{equation}
{U^\mu} _\nu =- u^\mu u_\nu,
\label{U}
\end{equation}
\begin{equation}
{h^\mu} _\nu ={\delta^\mu} _\nu +u^\mu u_\nu .
\label{h}
\end{equation}

\noindent Due to the existence of this singled out timelike direction $u$ a covariant 1+3 tensor decomposition of all geometrical objects of physical value can be made using $U$ and $h$.

We will denote a covariant derivative by $\nabla$ and the totally antisymmetric permutation tensor by $\eta^{\mu\nu\rho\sigma}$.

The well known kinematical fields associated with the timelike congruence $u$ are defined by

\begin{equation}
\dot{u}^\mu={h^\mu}_\nu u^\rho\nabla_\rho u^\nu,
\label{dotu}
\end{equation}
\begin{equation}
\Theta ={h^\mu}_\nu \nabla_\mu u^\nu,
\label{Theta}
\end{equation}
\begin{equation}
\sigma_{\mu\nu}=[{h^\rho}_{(\mu} {h^\sigma}_{\nu )}-\frac{1}{3}h_{\mu \nu}h^{\rho\sigma}](\nabla_\rho u_\sigma ),
\label{sigma}
\end{equation}
\begin{equation}
\omega_{\mu\nu}=-{h^\rho}_{[\mu} {h^\sigma}_{\nu ] }(\nabla_\rho u_\sigma ),
\label{omegatensor}
\end{equation}

\noindent where $\dot{u}^\mu$ is the acceleration vector, $\Theta$ the rate of expansion scalar, $\sigma_{\mu\nu}$ is the rate of shear tensor and $\omega_{\mu\nu}$ is the vorticity tensor. Note that $\sigma_{\mu\nu}$ is symmetric and tracefree and $\omega_{\mu\nu}$ is anti-symmetric. The magnitude of the rate of shear $\sigma$, the vorticity vector $\omega^{\mu}$ and the magnitude of the vorticity $\omega$ are defined as
\begin{equation}
\sigma^2=\frac{1}{2}\sigma_{\mu\nu}\sigma^{\mu\nu },
\label{sigma2}
\end{equation}
\begin{equation}
\omega^{\mu}=\frac{1}{2}\eta^{\mu\nu\rho\sigma}\omega_{\nu\rho}u_\sigma,
\label{omegavector}
\end{equation}
\begin{equation}
\omega^2=\frac{1}{2}\omega_{\mu\nu}\omega^{\mu\nu }=\omega_{\mu}\omega^{\mu}.
\label{omega2}
\end{equation}

The vector field $u$ is hypersurface forming if $\omega=0$.

In the orthonormal frame approach one chooses at each point of the spacetime manifold $(M,g)$ a set of four linearly independent 1-forms $e^a$ such that the line element is given by
\begin{equation}
ds^2= \eta_{ab}e^ae^b,
\label{lineelement}
\end{equation}

\noindent where $\eta_{ab}=diag[-1,1,1,1]$ is the constant Minkowskian frame metric. The vectors ${e_a}$ represent the dual basis.

In the 1+3 orthonormal frame formalism one aligns the timelike direction of the orthonormal frame with the tangent of the preferred timelike congruence ($e_0=u$).

\subsection{Commutators}

The commutator functions are defined by

\begin{equation}
[e_a,e_b]={\gamma^c}_{ab}e_c,
\label{comfunctions}
\end{equation}

\noindent where the frame vectors $e_a$ act as differential operators on geometrical objects.  The Ricci rotation coefficients ${\Gamma^a}_{bc}$ can be written in terms of the commutation functions as

\begin{equation}
{\gamma^a}_{bc}={\Gamma^a}_{cb}-{\Gamma^a}_{bc}.
\label{riccirotationcoef}
\end{equation}

The commutator functions with one or two indices equal to zero can be expressed in terms of the frame components of the kinematic quantities associated with the timelike congruence. The angular velocity $\Omega^a$ is defined by
\begin{equation}
\Omega^{a}=\frac{1}{2} \eta^{abcd}e_b\cdot \nabla_u({e}_c)u_d.
\label{Omega}
\end{equation}

The purely spatial components of the commutation functions ${\gamma^\alpha}_{\beta\gamma}$ can be decomposed \cite{ellismac}, depending on $a_\alpha$ and $n_{\alpha\beta}$, by means of

\begin{equation}
{\gamma^\alpha}_{\beta\gamma}=2a_{[\beta}{\delta^\alpha}_{\gamma ]}+\epsilon_{\beta\gamma\delta}n^{\delta\alpha},
\label{an}
\end{equation}

\noindent where $n_{\alpha\beta}$ is symmetric and $\epsilon_{\beta\gamma\delta}$ is the totally antisymmetric three-dimensional permutation tensor.

The commutators are given by expressions for ${\gamma^a}_{bc}$ and their 1+3 decomposition results in the following equations
\begin{eqnarray}
[e_0,e_\i ] &=&\dot{u}_\i e_0-[\frac{1}{3}\Theta{\delta^\j}_\i+{\sigma^\j}_\i+{\epsilon^\j}_{\i\k}(\omega^\k-\Omega^\k)]e_\j\label{com1}\\
{[e_\i ,e_\j ]} &=& -2\epsilon_{\i\j\k}\omega^\k e_0+[2a_{[\i}{\delta^\k}_{\j ]}
+\epsilon_{\i\j\m}n^{\m\k}]e_\k .
\label{com2}
\end{eqnarray}

\subsection{Curvature variables and their field equations}

The relation between the Riemann curvature tensor and the Ricci rotation coefficients is given by

\begin{equation}
{R^a}_{bcd}=e_c({\Gamma^a}_{bd})-e_d({\Gamma^a}_{bc})+{\Gamma^a}_{ec}{\Gamma^e}_{bd}-{\Gamma^a}_{ed}{\Gamma^e}_{bc}-{\Gamma^a}_{be}{\gamma^e}_{cd},
\label{Rabcd}
\end{equation}

\noindent and the 16 Jacobi identities become
\begin{equation}
{R^a}_{[bcd]}=0; \,\,\,R_{abcd}=R_{cdab}.
\label{Jacobiequations}
\end{equation}



The Einstein field equations with a non zero cosmological constant can be written as
\begin{equation}
{R}_{ab}-\frac{1}{2}R{g}_{ab}+\Lambda{g}_{ab}=T_{ab},
\label{EFQ}
\end{equation}

\noindent and the 1+3 decomposition of the energy-momentum-stress tensor $T_{ab}$ with respect to the timelike vector field $u$
in the fluid description of phenomena of a matter source is given by

\begin{equation}
{T}_{ab}=\mu u_au_b+2q_{(a}u_{b)}+ph_{ab}+\pi_{ab},
\label{energytensor}
\end{equation}

\noindent where $\mu$ denotes the total energy density scalar, $p$ denotes the isotropic pressure scalar, $q^a$ denotes the energy current density vector, and $\pi_{ab}$ denotes the anisotropic pressure tensor. Note that
\begin{equation}
{q}_{a}u^a=0;\,\,\, \pi_{ab}u^b=0;\,\,\, {\pi^a}_{a}=0; \,\,\, \pi_{ab}=\pi_{ba}.
\label{Tab1}
\end{equation}

The matter fields need to satisfy an appropriate thermodynamical equation of state in order to describe the physics of the underlying fluid spacetime geometry under consideration.

The completely tracefree Weyl conformal curvature tensor is defined by
\begin{equation}
{C^{ab}}_{cd}={R^{ab}}_{cd}-2{\delta^{[a}}_{[c}{R^{b]}}_{d]}+\frac{1}{3}R{\delta^{a}}_{[c}{\delta^{b}}_{d]}.
\label{Weyl}
\end{equation}

When there is a singled out timelike direction $u$, it is convenient to decompose the Weyl tensor into its electric part given by
\begin{equation}
E_{ab}=C_{cedf}{h^c}_{a}u^e{h^d}_{b}u^{f},
\label{Eab}
\end{equation}

\noindent and magnetic part

\begin{equation}
H_{ab}={^*}C_{cedf}{h^c}_{a}u^e{h^d}_{b}u^{f},
\label{Hab}
\end{equation}

\noindent where ${^*}C_{abcd}$ is the dual of $C_{abcd}$ defined by

\begin{equation}
{^*}C_{abcd}=\frac{1}{2}{\eta_{ab}}^{ef}C_{efcd}.
\label{Weyldual}
\end{equation}

The electric and magnetic tensors are symmetric and tracefree and satisfy $E_{ab}u^b=H_{ab}u^b=0$ and lead to the expression

\begin{eqnarray}
&&{C^{ab}}_{cd}=[4{\delta^{[a}}_{e}{\delta^{b]}}_{f}{\delta^{g}}_{[c}{\delta^{h}}_{d]}-{\eta^{ab}}_{ef}{\eta^{gh}}_{cd}]{E^{e}}_{g}u^fu_ h\nonumber\\
&&- 2[{\eta^{ab}}_{ef}{\delta^{g}}_{[c}{\delta^{h}}_{d]}+{\delta^{[a}}_{e}{\delta^{b]}}_{f}{\eta^{gh}}_{cd}]{H^{e}}_{g}u^fu_ h.
\label{Weyl1}
\end{eqnarray}

Considering equation (\ref{Weyl}) and using (\ref{EFQ}), (\ref{energytensor}) and (\ref{Weyl1}) we obtain the following expression for the Riemann curvature tensor

\begin{eqnarray}
&&{R^{ab}}_{cd}=[4{\delta^{[a}}_{e}{\delta^{b]}}_{f}{\delta^{g}}_{[c}{\delta^{h}}_{d]}-{\eta^{ab}}_{ef}{\eta^{gh}}_{cd}]{E^{e}}_{g}u^fu_ h\nonumber\\
&&- 2[{\eta^{ab}}_{ef}{\delta^{g}}_{[c}{\delta^{h}}_{d]}+{\delta^{[a}}_{e}{\delta^{b]}}_{f}{\eta^{gh}}_{cd}]{H^{e}}_{g}u^fu_ h\nonumber\\
&&+2{\delta^{[a}}_{[c}[(\mu +p)u^{b]}u_{d]}+q^{b]}u_{d]}+u^{b]}q_{d]}+{\pi^{b]}}_{d]}]\nonumber\\
&&+\frac{2}{3}(\mu +\Lambda){\delta^{a}}_{[c}{\delta^{b}}_{d]}.
\label{curvature}
\end{eqnarray}

Inserting expression (\ref{Rabcd}) for the Riemann tensor into the left hand side of (\ref{curvature}) one obtains the 10 Einstein field equations, the 16 Jacobi identities and expressions for $E_{\alpha\beta}$ and $H_{\alpha\beta}$ in terms of the basis variables

$$\{ \Theta,\,\,\dot{u}_\alpha,\,\,\, \sigma_{\alpha\beta}, \,\,\, \omega_\alpha,\,\,\, \Omega_\alpha, \,\,\, a_\alpha,\,\,\, n_{\alpha\beta},\,\,\, \mu,\,\,\, p, \,\,\, q_\alpha,\,\,\, \pi_{\alpha\beta}\},$$

\noindent and their $e_0$ and $e_\alpha$ frame derivatives.

\bigskip

{\bf Einstein field equations}

\begin{eqnarray}
&&e_0(\Theta)=-\frac{1}{3}\Theta^2+(e_\alpha +\dot{u}_\alpha -2a_\alpha )(\dot{u}^\alpha )-2\sigma^2+2\omega^2\nonumber\\
&&-\frac{1}{2}(\mu +3p)+\Lambda,\label{field1}\\
&&e_0(\sigma^{\i\j})=-\Theta\sigma^{\i\j}+(\delta^{\gamma (\alpha}e_\gamma +\dot{u}^{( \alpha} +{a}^{( \alpha})(\dot{u}^{\beta )})
+2\omega^{( \alpha}\Omega^{\beta )}+\pi^{\i\j}\nonumber\\
&&- ^\star{S}^{\i\j}-\frac{1}{3}\delta^{\i\j}[(e_\gamma+\dot{u}_\gamma +a_\gamma)(\dot{u}^\gamma )+2\omega_\gamma\Omega^\gamma]\nonumber\\
&&+\epsilon^{\gamma\delta (\alpha}[2\Omega_\gamma{\sigma^{\beta )}}_\delta -{n^{\beta )}}_\gamma\dot{u}_\delta],\label{field2}\\
&&0=\mu-\frac{1}{3}\Theta^2+\sigma^2-\omega^2-2\omega_\i\Omega^\i-\frac{1}{2} {^\star{R}}+\Lambda,\label{field3}\\
&&0=(e_\j-3a_\j)(\sigma^{\i\j})-\frac{2}{3}\delta^{\i\j}e_\j(\Theta)+{n^\i}_\j\omega^\j+q^\i\nonumber\\
&&-\epsilon^{\i\j\gamma}[(e_\j+2\dot{u}_\j-a_\j)(\omega_\gamma)+n_{\j\delta}{\sigma^\delta}_\gamma],\label{field4}
\end{eqnarray}

\noindent

where

\begin{eqnarray}
&&^\star{S}_{\i\j}=e_{(\alpha} (a_{\beta )})+b_{\i\j}-\frac{1}{3}\delta_{\i\j}[e_\gamma(a^\gamma)+{b^\gamma}_\gamma ]\nonumber\\
&&-{\epsilon^{\gamma\delta}}_{(\i}(e_{|\gamma|}-2a_{|\gamma|})
(n_{\j )\delta}),\label{field5}\\
&&^\star{R}=2(2e_\i-3a_\i)(a^\i)-\frac{1}{2}{b^\i}_\i,\label{field6}\\
&&b_{\i\j}=2n_{\i\gamma}{n^\gamma}_\j-{n^\gamma}_\gamma n_{\i\j}.\label{field7}
\end{eqnarray}

\bigskip

{\bf Jacobi identities}

\begin{eqnarray}
&&e_0(a^\i)=-\frac{1}{3}(\delta^{\i\j}e_\j+\dot{u}^\i+a^\i)(\Theta)+\frac{1}{2}(e_\j+\dot{u}_\j-2a_\j)(\sigma^{\i\j})\nonumber\\
&&-\frac{1}{2}\epsilon^{\i\j\k}(e_\j+\dot{u}_\j-2a_\j)(\omega_\k-\Omega_\k),\label{jacobi1}\\
&&e_0(n^{\i\j})=-\frac{1}{3}\Theta n^{\i\j}-(\delta^{\k(\i}e_\k+\dot{u}^{(\i})(\omega^{\j )}-\Omega^{\j )})+2{\sigma^{(\i}}_\k n^{\j )\k}\nonumber\\
&&+\delta^{\i\j}(e_\k+\dot{u}_\k)(\omega^{\k}-\Omega^{\k})-\epsilon^{\k\m(\i}[(e_\k+\dot{u}_\k)({\sigma^{\j )}}_\m)\nonumber\\
&&-2{n^{\j )}}_\k(\omega_\m-\Omega_\m)],\label{jacobi2}\\
&&e_0(\omega^{\i})=-\frac{2}{3}\Theta\omega^\i+{\sigma^\i}_\j\omega^\j+\frac{1}{2}{n^\i}_\j\dot{u}^\j-\epsilon^{\i \j \k}[\frac{1}{2}(e_\j-a_\j)(\dot{u}_\k)\nonumber\\
&&+ \omega_\j\Omega_\k ],\label{jacobi3}\\
&&0=(e_\j-2a_\j)(n^{\i \j})-\frac{2}{3}\Theta\omega^\i-2{\sigma^\i}_\j\omega^\j+\epsilon^{\i \j \k}[e_\j(a_\k)\nonumber\\
&&+2\omega_\j\Omega_\k],\label{jacobi4}\\
&&0=(e_\i-\dot{u}_\i-2a_\i)(\omega^\i).\label{jacobi5}
\end{eqnarray}

The Bianchi identities are given by

\begin{equation}
\nabla_{[a}R_{bc]de}=0\Longleftrightarrow -\nabla_{[a}[{R^c}_{b]}-\frac{1}{6}{\delta^c}_{b]}R]=-\nabla_{[a}[{T^c}_{b]}-\frac{1}{3}{\delta^c}_{b]}T],
\label{BI}
\end{equation}

\noindent and can be expressed in terms of $E_{\alpha\beta}$ and $H_{\alpha\beta}$ and the 1+3 basis variables associated with $T_{ab}$

\bigskip

{\bf Bianchi identities}

\begin{eqnarray}
&&e_0(\mu)=-(\mu+p)\Theta-(e_\i+2\dot{u}_\i-2a_\i)(q^\i)-\sigma_{\i \j}\pi^{\i\j},\label{bianchi1}\\
&&e_0(q^\i)=-\frac{4}{3}\Theta q^\i-\delta^{\i\j}e_\j(p)-(\mu+p)\dot{u}^\i-(e_\j+\dot{u}_\j-3a_\j)(\pi^{\i\j})\nonumber\\
&&-{\sigma^\i}_\j q^\j +\epsilon^{\i\j\k}[(\omega_\j +\Omega_\j)q_\k +n_{\j \m}{\pi^\m}_\k],\label{bianchi2}\\
&&e_0(E^{\i\j}+\frac{1}{2}\pi^{\i \j})=-\frac{1}{2}(\mu+p)\sigma^{\i \j}-\Theta(E^{\i\j}+\frac{1}{6}\pi^{\i \j})\nonumber\\
&&-\frac{1}{2}(\delta^{\k (\i}e_\k
+2\dot{u}^{(\i}+a^{(\i})(q^{\j )})+3{\sigma^{(\alpha}}_\gamma (E^{\j )\gamma}-\frac{1}{6}\pi^{\j )\gamma})+\frac{1}{2}{n^\gamma}_\gamma H^{\i\j}\nonumber\\
&&+\frac{1}{3}\delta^{\i\j}\left[\frac{1}{2}(e_\gamma+2\dot{u}_\gamma+a_\gamma )(q^\k)-3\sigma_{\k\delta}(E^{\k\delta}-\frac{1}{6}\pi^{\k\delta})+3n_{\k\delta}H^{\k\delta}\right]\nonumber\\
&&+\epsilon^{\k\delta (\i}\left[(e_\k +2\dot{u}_\gamma-a_\gamma )({H^{\j )}}_\delta )-(\omega_\k -2\Omega_\k )({E^{\j )}}_\delta+\frac{1}{2}{\pi^{\j )}}_\delta )+\frac{1}{2}{n^{\j )}}_\k q_\delta\right]\nonumber\\
&&-3{n^{(\alpha}}_\gamma H^{\beta )\gamma}, \label{bianchi3}\\
&&e_0(H^{\i\j})= -\Theta H^{\i\j}+3{\sigma^{(\i }}_\k H^{\j )\k}-\frac{3}{2}\omega^{(\alpha}q^{\j )}-\frac{1}{2}{n^\k}_\k(E^{\i\j}-\frac{1}{2}\pi^{\i\j})\nonumber\\
&&+3{n^{(\i }}_\k (E^{\beta )\k}-\frac{1}{2}\pi^{\beta )\k}) -\delta^{\i\j}\left[\sigma_{\k\delta}H^{\k\delta}-\frac{1}{2}\omega_\k q^\k +n_{\k\delta}(E^{\k\delta}-\frac{1}{2}\pi^{\k\delta})\right]\nonumber\\
&&-\epsilon^{\k\delta (\alpha}[(e_\k -a_\k )({E^{\j ) }}_\delta -\frac{1}{2}{\pi^{\j )}}_\delta )+2\dot{u}_\gamma {E^{\j )}}_\delta -\frac{1}{2}{\sigma^{\j )}}_\k q_\delta\nonumber\\
&&+ (\omega_\gamma -2\Omega_\gamma ){H^{\j )}}_\delta],\label{bianchi4}\\
&&0= (e_\j -3a_\j )(E^{\i\j}+\frac{1}{2}\pi^{\i\j})-\frac{1}{3}\delta^{\i\j}e_\j (\mu )+\frac{1}{3}\Theta q^\i-\frac{1}{2}{\sigma^{\i }}_\j q^\j \nonumber\\
&&+3\omega_\j H^{\i\j}-\epsilon^{\i\j\k}[\sigma_{\j\delta}{H^{\delta }}_\k + \frac{3}{2}\omega_\j q_\k + n_{\j\delta}({E^{\delta}}_\gamma +\frac{1}{2}{\pi^{\delta}}_\gamma )],\label{bianchi5}\\
&&0 = (e_\j -3a_\j )(H^{\i\j})-(\mu +p)\omega^\i -3\omega_\j(E^{\i\j}-\frac{1}{6}\pi^{\i\j})-\frac{1}{2}{n^\alpha}_\j q^\j  \nonumber\\
&&+\epsilon^{\i\j\k}[\frac{1}{2}(e_\j -a_\j)(q_\gamma )+\sigma_{\j\delta}({E^{\delta }}_\gamma +\frac{1}{2}{\pi^{\delta }}_\gamma )- n_{\j\delta}{H^{\delta }}_\gamma ].\label{bianchi6}
\end{eqnarray}

For a prescribed set of matter equations of state, the commutators (\ref{com1})-(\ref{com2}), the Einstein field equations (\ref{field1})-(\ref{field4}), the Jacobi identities (\ref{jacobi1})-(\ref{jacobi5}) and the Bianchi identities (\ref{bianchi1})-(\ref{bianchi6}) form a complete set of tetrad relations, which can be used to determine particular classes of solutions.

\section{Relasticity}
\label{sec:relasticity}

In this section we will use the formulation given in \cite{karl1} for the theory of relativistic elasticity .
The theory of relativistic elasticity is based on a configuration mapping $\psi: M \rightarrow X$ from the spacetime $M$ to the three-dimensional material space $X$, which represents the collection of particles of the material. The material coordinates will be denoted by $y^A,$ $A=1,2,3,$ and let $x^\mu$ represent the coordinates in $M$. The material space $X$ is equipped with a particle density form $n_{ABC}=n_{[ABC]}.$ The mapping $\psi$ describes the configuration of the material and gives rise to a rank three matrix
$ \left( y^A_{\; \mu}\right)_p, \,\, p\in M,\;\;\;  y^A_{\; \mu}=\frac{\partial y^A}{\partial x^\mu},$ which is called the relativistic deformation gradient. The velocity field of matter $u^\mu$, a future oriented, timelike unit vector field, which spans the one-dimensional Kernel of the relativistic deformation gradient is defined by the conditions $y^A_{\; \mu} u^\mu = 0, \quad u^\mu u_\mu = -1.$
The pulled-back density form
\begin{equation}
n_{\mu\nu\rho}=\psi^{\ast}n_{ABC}=y^A_{\; \mu}y^B_{\; \nu}y^C_{\; \rho}n_{ABC}
\end{equation}
can be used to define the flowline tangential particle current
\begin{equation}
n^{\mu}=\frac{1}{3!}\epsilon^{\mu\nu\rho\sigma}n_{\nu\rho\sigma},
\end{equation}
 $\epsilon^{\mu\nu\rho\sigma}$ being the spacetime volume form associated with $g_{\mu\nu}.$ The particle density $n$ is such that $n^{\mu}=nu^{\mu},$ $n=\sqrt{-n^{\mu} n_{\mu}},$ $\nabla_{\mu}n^{\mu}=0$ and satisfies
\begin{equation}
n_{\mu\nu\rho}=n\epsilon_{\mu\nu\rho},
\end{equation}
where $\epsilon_{\mu\nu\rho}=\epsilon_{\mu\nu\rho\sigma}u^{\sigma}$ is the spatial volume form.
Let $\eta_{AB}$ be an $n$-dependent tensor defined in $X$ having $\epsilon_{ABC}$ as its volume form 
 The relation between $\epsilon_{ABC}$ and the particle density form $n_{ABC}$ is given by $n_{ABC}=n\epsilon_{ABC}$. The pulled-back tensor $\eta_{\mu\nu}=\psi^{*}\eta_{AB}$
can be used to construct the strain tensor
\begin{equation}
s_{\mu\nu} = \frac12(h_{\mu\nu} - \eta_{\mu\nu}),
\label{straintensor}
\end{equation}
where $h_{\mu\nu} = g_{\mu\nu} + u_\mu u_\nu$. If $s_{\mu\nu}=0,$ the material is said to be in a locally unsheared state for some particle density $n$.

The energy density can be defined by $\mu=n\epsilon,$ where $\epsilon$ represents the energy per particle. We assume that $\epsilon$ has a minimum value $\tilde{\epsilon}$ under variations of $g^{AB}=\Psi_{\ast}g^{ab}$ such that the particle density $n$ is held fixed. This state is called the unsheared state and in this case, when $\epsilon=\tilde{\epsilon}$, the tensor $\eta_{AB}$ is such that $g^{AC}\eta_{CB}=\delta^{A}_{B}$.

The stress-energy tensor for elastic matter is given by

\begin{equation}
T_{\mu\nu}=\mu u_\mu u_\nu +p_{\mu\nu},
\label{Telasticity}
\end{equation}

By specifying $\epsilon$ as a function of $n$ and ${\eta^\mu}_\nu$, the pressure tensor can be written as

\begin{equation}
p_{\mu\nu}=ph_{\mu\nu}+\pi_{\mu\nu}
\label{pelasticity}
\end{equation}

\noindent where

\begin{equation}
p=n^2\frac{\partial\epsilon}{\partial n},\,\,\,\,\, \pi_{\mu\nu}=2n\eta_{\rho (\mu}\frac{\partial\epsilon}{\partial {\eta^{\nu )}}_\rho}.
\label{pelasticity1}
\end{equation}

This shows that the dependence of $\epsilon$ on the particle density $n$ is directly related to the isotropic pressure $p$, while the dependence on ${\eta^\mu}_\nu$ is directly related to the anisotropic pressure tensor $\pi_{\mu\nu}$. The energy density is a function of only the number density if and only if the matter source corresponds to a perfect fluid. In this last case the equation of state is $\mu =n\epsilon$, $p=n^2 \frac{d\epsilon}{dn}$ where $\epsilon$ coincides with the unsheared quantity $\tilde{\epsilon}$ which is the minimum energy per particle at fixed particle density.

In the next section we will assume an equation of state where the energy per particle depends on only one invariant of ${\eta^\mu}_\nu$, namely the shear scalar given by

\begin{equation}
s^2=\frac{1}{36}\left[\left(\eta^{\mu}_{\hspace{0.13cm}\mu}\right)^3-\eta^{\mu}_{\hspace{0.13cm}\nu}\eta^{\nu}_{\hspace{0.13cm}\rho}
\eta^{\rho}_{\hspace{0.13cm}\mu}-24\right].
\label{shearscalar}
\end{equation}

An equation of state that is compatible with the stress energy tensor (\ref{Telasticity}) can be written in the form

\begin{equation}
\mu=\tilde{\mu} +\tilde{\rho}s^2,
\label{equationstate1}
\end{equation}

\noindent where $\tilde{\mu}$ is the unsheared energy density and $\tilde{\rho}$ is the modulus of rigidity. All unsheared quantities are functions of $n$ only.

For this class of equations of state the pressure tensor $p$ takes the form
\begin{eqnarray}
&&p=\tilde{p} +(\tilde{\Omega}-1)\tilde{\rho}s^2 , \,\,\,\, \tilde{p}=n^2\frac{d\tilde{\epsilon}}{dn},\,\,\,\, \tilde{\Omega}=\frac{n}{\tilde{\rho}}\frac{d\tilde{\rho}}{dn},\label{equationstate2}\\
&&\pi_{\mu\nu}=\frac{1}{6}\tilde{\rho}\left[({\eta^\rho}_\rho )^2\eta_{\langle\mu\nu \rangle}-\eta^{\rho\sigma}\eta_{\rho \langle\mu}\eta_{\nu\rangle\sigma}\right],
\label{equationstate3}
\end{eqnarray}

\noindent where the angle brackets $\langle \cdot\cdot\rangle$ of an index pair denote the symmetric and traceless part (with respect to $h_{\mu\nu}$) of the pair, e.g. for a tensor $t_{\mu\nu}$ one has
\begin{equation}
t_{\langle \mu\nu\rangle}=t_{(\mu\nu)}-\frac{1}{3}t^{\gamma}_{\hspace{0.13cm}\gamma}h_{\mu\nu}.
\end{equation}

\section{Conformally flat metrics with elastic stress energy tensor}
\label{sec:conflat}

It is well known that the Weyl tensor $C_{\mu\nu\rho\sigma}$ vanishes iff the spacetime is conformally flat. By definition, the metric of a conformally flat spacetime can be written as

\begin{equation}
ds^2= F^2(t,x,y,z)(-dt^2+dx^2+dy^2+dz^2).
\label{conflatmetric}
\end{equation}

All conformally flat solutions with a perfect fluid, an electromagnetic field, or a pure radiation field are known.

\subsection{A special case}\label{special}

Here we study the special case of conformally flat spacetimes with elastic energy-momentum tensor (\ref{Telasticity}), with an anisotropic traceless pressure tensor $\pi_{\mu\nu}$ that is of the form

\begin{equation}
\pi_{33}=-2\pi_{11}=-2\pi_{22},
\label{anispressure}
\end{equation}

\noindent with all other components being null. We will also assume that the conformal factor $F$ and the component $\pi_{11}$ are functions of only one spatial coordinate, i.e. $F=F(z)$; $\pi_{11}=\pi_{11}(z)$.

We can relate the Ricci tensor components to the energy density, the isotropic pressure and anisotropic pressure via (\ref{Telasticity}) and (\ref{EFQ}). We choose to use the equivalent formulation in terms of Ricci spinor components given its more compact form. In doing so we obtain

\begin{eqnarray}
&&\Phi_{00^\prime}=\Phi_{22^\prime} = \frac{1}{4}(\mu+p+\pi_{33})\label{ricci1}\\
&&\Phi_{11^\prime}= \frac{1}{8}(\mu+p+\pi_{11}+\pi_{22}-\pi_{33})\label{ricci2}\\
&&\Phi_{12^\prime}=\overline{\Phi}_{21^\prime}=-\Phi_{01^\prime}=-\overline{\Phi}_{10^\prime} = \frac{1}{4}(\pi_{13}-i\pi_{23})\label{ricci3}\\
&&\Phi_{02^\prime}=\overline{\Phi}_{20^\prime} = \frac{1}{4}(\pi_{11}-\pi_{22}-i2\pi_{12})\label{ricci4}\\
&&\frac{1}{24}R =\frac{1}{24}(\mu-3p)-\frac{1}{6}\Lambda.\label{ricciscal}
\end{eqnarray}

 \noindent On the lefthand side we have the Ricci spinor components and on the righthand side we have the 1+3 components of the symmetric anisotropic pressure tensor $\pi_{ab}$, the energy density scalar $\mu$ and isotropic pressure scalar $p$.

In the case under study the above relations simplify considerably as is shown below

\begin{eqnarray}
&&\Phi_{00^\prime}=\Phi_{22^\prime} = \frac{1}{4}(\mu+p-2\pi_{11}),\label{ricci1s}\\
&&\Phi_{11^\prime}= \frac{1}{8}(\mu+p+4\pi_{11}),\label{ricci2s}\\
&&\frac{1}{24}R =\frac{1}{24}(\mu-3p)-\frac{1}{6}\Lambda,\label{ricciscals}
\end{eqnarray}

\noindent with all Ricci spinor components being null.

If we consider the coordinates $(t,x,y,z)$ and the 1+3 basis

\begin{eqnarray}
&&{e_0}_\mu=u_\mu=\left( F(z),0,0,0\right);\,\,\, {e_1}_\mu=\left( 0,F(z),0,0\right);\nonumber\\
&&{e_2}_\mu=\left( 0,0,F(z),0\right);\,\,\, {e_3}_\mu=\left( 0,0,0,F(z)\right),\label{basis}
\end{eqnarray}

\noindent then the Ricci spinor components in this basis are:

\begin{eqnarray}
&&\Phi_{00^\prime}=\Phi_{22^\prime} = -2\Phi_{11^\prime}=\frac{F_{zz}F+2F_z^2}{2F^4}\label{ricci1sbasis}\\
&&\frac{1}{24}R =-\frac{F_{zz}}{4F^3}.\label{ricciscalsbasis}
\end{eqnarray}

So that in this basis the fact that $\Phi_{22^\prime} = -2\Phi_{11^\prime}$ implies by (\ref{ricci1s}) and (\ref{ricci2s}) that $\mu +p+\pi_{11}=0$.

\subsection{1+3 treatment of the special case}

In the particular case described in this section, we assume that the pressure $p$, the energy density $\mu$ and the only non null independent anisotropic pressure component $\pi_{11}$ are functions of one spatial coordinate $z$ . Therefore, the following conditions must hold, in order for the commutators (\ref{com1})-(\ref{com2}) acting on these quantities to be satisfied

\begin{eqnarray}
&&\Omega_1=\omega_1-\sigma_{23}\label{C1}\\
&&\Omega_2=\omega_2+\sigma_{13}\label{C2}\\
&&\Theta =3(\sigma_{11}+\sigma_{22})\label{C3}\\
&&n_{23}=a_1\label{C4}\\
&&n_{13}=-a_2\label{C5}\\
&&n_{33}=0,\label{C6}
\end{eqnarray}

\noindent otherwise $\pi_{11}$, $p$ and $\mu$ must all be constant.

If we calculate the Bianchi identities (\ref{bianchi1})-(\ref{bianchi6}), considering that (\ref{C1})-(\ref{C6}), obtained from the commutators, must hold, we get the following equations for the 1+3 directional derivative $e_3$ acting on $p$, $\mu$ and $\pi_{11}$ as well as conditions involving $p,\,\,\pi_{11},\,\, \mu ,\,\,n_{\alpha\beta},\,\, a_\alpha,\,\, \dot{u}_\alpha,\,\, \omega_\alpha,\,\, \sigma_{\alpha\beta},\,\, \Omega_\alpha ,\,\, \Theta$

\begin{eqnarray}
&&\mu +p+\pi_{11}=0\,\,\, or\,\,\, ({\sigma}_{11}= {\sigma}_{22}={\sigma}_{12}=\dot{u}_1=\dot{u}_2=\omega_3=\nonumber\\
&&\omega_2+{\sigma}_{13}=\omega_1-{\sigma}_{23}=0)\label{b1}\\
&&n_{12}=n_{11}-n_{22}=0\label{b2}\\
&&2(\mu +p)\omega_2-(\omega_2+3{\sigma}_{13}){\pi}_{11}=0\label{b3}\\
&&2(\mu +p)\omega_1-(\omega_1-3{\sigma}_{23}){\pi}_{11}=0\label{b4}\\
&&2e_3(\pi_{11})-e_3(p)=(\mu +p)\dot{u}_3+2(3a_3-\dot{u}_3)\pi_{11}\label{b5}\\
&&3e_3(\pi_{11})+e_3(\mu)=9a_3\pi_{11}.\label{b6}
\end{eqnarray}

In what follows we outline the cases that arise from condition (\ref{b1}).

\bigskip

{\bf{Case A}}: $\mu+p+\pi_{11}=0$

\bigskip

Equations (\ref{C1})-(\ref{C6}) obtained from the commutators and (\ref{b1})-(\ref{b6}) resulting from the Bianchi identities give $\Omega_1=\Omega_2=n_{33}=n_{12}=n_{11}-n_{22}=n_{13}+a_2=n_{23}-a_1=a_3+\dot{u}_3=\omega_1-\sigma_{23}
=\omega_2+\sigma_{13}=0$ along with the ODE in the variables $p$ and $\mu$

\begin{equation}
2e_3(\mu )+3e_3(p)=-9(\mu +p)\dot{u}_3.\label{ONE1A}
\end{equation}

\bigskip

Within this case we will analyse the particular solution with $n_{13}=a_2=n_{23}=a_1=\dot{u}_1=\dot{u}_2=\omega_1=\omega_2=\sigma_{13}=\sigma_{23}=0$, which is described below as case {\bf A*}.

\bigskip

{\bf{Case A*}}: $\mu+p+\pi_{11}=a_3+\dot{u}_3=n_{11}-n_{22}=n_{12}=n_{33}=n_{13}=n_{23}=\Omega_1=\Omega_2=a_2=a_1=
\dot{u}_1=\dot{u}_2=\omega_1=\omega_2=\sigma_{13}=\sigma_{23}=0$.

\bigskip

The commutators give no extra information since they reduce to identities. The Einstein equations (\ref{field1})-(\ref{field7}) and the Jacobi identities (\ref{jacobi1})-(\ref{jacobi5}) in this case give

\bigskip

\begin{eqnarray}
&&(\sigma_{11}+\sigma_{22})\omega_3=0\label{JE1}\\
&&(\sigma_{11}-\sigma_{22})\Omega_3=3(\sigma_{11}+\sigma_{22})\sigma_{12}\label{JE2}\\
&&\sigma_{22}^2-\sigma_{11}^2=\frac{4}{3}\sigma_{12}\Omega_3\label{JE3}\\
&&(\sigma_{22}-\sigma_{11})^2=4\omega_3(\omega_3+\Omega_3)-4\sigma_{12}^2\label{JE4}\\
&&\mu+\frac{3}{2}p=-4(\sigma_{22}^2+\sigma_{11}^2)-\frac{11}{2}\sigma_{11}\sigma_{22}+\frac{5}{2}\omega_3^2+\omega_3\Omega_3\nonumber\\
&&+\frac{1}{2}\Lambda+\frac{3}{2}a_3^2 -\frac{5}{2}\sigma_{12}^2\label{JE5}\\
&&e_3(\omega_3)=-\dot{u}_3\omega_3\label{JE6}\\
&&e_3(\Omega_{3})=-\dot{u}_3\Omega_{3}\label{JE7}\\
&&e_3(\sigma_{12})=(\sigma_{22}-\sigma_{11})n_{11}-\dot{u}_3\sigma_{12}\label{JE8}\\
&&e_3(\sigma_{11})=-\dot{u}_3\sigma_{11}+2n_{11}\sigma_{12}\label{JE9}\\
&&e_3(\sigma_{22})=-\dot{u}_3\sigma_{22}-2n_{11}\sigma_{12}\label{JE10}\\
&&e_3(\dot{u}_3)=5(\sigma_{11}^2+\sigma_{22}^2)+8\sigma_{11}\sigma_{22}-3\dot{u}_3^2-2\omega_3^2\nonumber\\
&&-\Lambda+\frac{1}{2}(\mu +3p)+2\sigma_{12}^2.\label{JE11}
\end{eqnarray}

The conditions in (\ref{JE1})-(\ref{JE4}) give rise to the following three subcases:

\bigskip

{\bf Case A1*}: $\Omega_{\alpha}=\sigma_{22}+\sigma_{11}=\sigma_{11}^2+\sigma_{12}^2-\omega_3^2=0$

\bigskip

   The solutions here have zero angular velocity $\Omega_{\alpha}=0$, rotation and shear components satisfying $\sigma_{11}^2=\omega_3^2-\sigma_{12}^2$, $\sigma_{22}=-\sigma_{11}$,  and a non zero acceleration term where $a_3=-\dot{u}_3$. They are characterised by the following equations

\begin{eqnarray}
&&e_3(\dot{u}_3)=-\frac{3}{2}\dot{u}_3^2-\frac{1}{2}\Lambda-\frac{1}{2}\mu\label{A1*1}\\
&&e_3(\sigma_{11})=-\dot{u}_3\sigma_{11}+2n_{11}\sigma_{12}\label{A1*2}\\
&&e_3(\sigma_{12})=-\dot{u}_3\sigma_{12}-2n_{11}\sigma_{11}\label{A1*3}\\
&&\mu +\frac{3}{2}p=\frac{1}{2}\Lambda+\frac{3}{2}\dot{u}_3^2.\label{A1*4}
\end{eqnarray}

\bigskip

If we introduce the simple equation of state (\ref{equationstate1})-(\ref{equationstate3}) in the previous equations we obtain the following ODE system in the variables $\sigma_{11}$, $\sigma_{12}$  and $\dot{u}_3$

\begin{eqnarray}
&&e_3(\dot{u}_3)=-\frac{3}{2}\dot{u}_3^2-\frac{1}{2}\Lambda-\frac{1}{2}\tilde{\mu}-\frac{1}{2}\tilde{\rho}s^2\label{A1*1es}\\
&&e_3(\sigma_{11})=-\dot{u}_3\sigma_{11}+2n_{11}\sigma_{12}\label{A1*2es}\\
&&e_3(\sigma_{12})=-\dot{u}_3\sigma_{12}-2n_{11}\sigma_{11},\label{A1*3es}
\end{eqnarray}

\noindent along with the conditions on the unsheared quantities $\tilde{\mu}$, $\tilde{p}$, $\tilde{\Omega}$, $\tilde{\rho}$ and pull back material metric $\eta_{\alpha\beta}$

\begin{eqnarray}
&&\tilde{\mu}+\frac{3}{2}\tilde{p}+\frac{1}{2}(3\tilde{\Omega}-1)\tilde{\rho}s^2
=\frac{1}{2}\Lambda+\frac{3}{2}\dot{u}_3^2\label{A1*4es}\\
&&\pi_{11}=-(\tilde{\mu} +\tilde{p}+\tilde{\Omega}\tilde{\rho}s^2)=\frac{\tilde{\rho}}{6F^2}\left[({\eta^\rho}_\rho )^2\eta_{\langle 11\rangle}-\eta^{\rho\sigma}
\eta_{\rho \langle 1}\eta_{1\rangle\sigma}\right]\label{A1*5es}\\
&&\tilde{\rho}\left[({\eta^\rho}_\rho )^2\eta_{\langle 22 \rangle}-\eta^{\rho\sigma}\eta_{\rho \langle 2}\eta_{2\rangle\sigma}\right]=
\tilde{\rho}\left[({\eta^\rho}_\rho )^2\eta_{\langle11 \rangle}-\eta^{\rho\sigma}\eta_{\rho \langle 1}\eta_{1\rangle\sigma}\right],\label{A1*6es}
\end{eqnarray}

\noindent where $\langle11\rangle$, $\langle 22\rangle$ and $\langle 33\rangle$ refer to spatial basis components in the basis (\ref{basis}).

In fact, the tensorial equation (\ref{equationstate3}) gives five independent scalar relations. Assuming that $\eta_{\mu\nu}$ has the same symmetries as the metric $g_{\mu\nu}$, the three equations corresponding to $\langle12\rangle$, $\langle13\rangle$, $\langle23\rangle$ that do not appear above are trivially satisfied.

Integration of this system, using the coordinates and basis indicated in (\ref{basis}), shows that for a non null cosmological constant the only existing solution is a perfect fluid ($\pi_{11}=0$) with equation of state $\mu =-p$, the conformal factor taking the form $F(z)=\frac{1}{\alpha z+\beta}$, where $\alpha$ and $\beta$ are real constants. However when $\Lambda =0$ the system allows for a solution with $\pi_{11}\neq 0$

\begin{eqnarray}
&&n_{11}=0\\
&&\dot{u}_3=\frac{F_z}{F^2}\label{solA11}\\
&&\sigma_{11}=\sigma_{12}=-\sigma_{22}=\pm \frac{\sqrt{2}}{2}\omega_3=\frac{C}{F}\label{solA12}\\
&&\tilde{\mu}+\tilde{\rho}s^2=\frac{-2F_{zz}F+F_z^2}{F^4}\label{solA13}\\
&& \tilde{p}+(\tilde{\Omega}-1)\tilde{\rho}s^2=\frac{4F_{zz}F+F_z^2}{3F^4} \label{solA14}\\
&&\tilde{\rho}\left[({\eta^\rho}_\rho )^2\eta_{\langle11\rangle}-\eta^{\rho\sigma}
\eta_{\rho \langle1}\eta_{1\rangle\sigma}\right]=\frac{4F_{zz}F-8F_z^2}{F^2}\label{solA15}\\
&&\tilde{\rho}\left[({\eta^\rho}_\rho )^2\eta_{\langle22\rangle}-\eta^{\rho\sigma}
\eta_{\rho \langle2}\eta_{2\rangle\sigma}\right]=\tilde{\rho}\left[({\eta^\rho}_\rho )^2\eta_{\langle11\rangle}-\eta^{\rho\sigma}
\eta_{\rho \langle1}\eta_{1\rangle\sigma}\right],\label{solA16}
\end{eqnarray}

\noindent where $F=F(z)$ is a real $C^2$ non null function, and $C$ is a real arbitrary constant. Note that the shearing scalar $s^2$ is given in terms of $\eta_{\mu\nu}$ by (\ref{equationstate3}).

In the absence of elasticity ($s^2=\tilde{\rho}=0$) the previous system produces a perfect fluid solution with $\mu =-p$ and $F(z)$ being the solution of the ODE $\frac{-F_{zz}F+2F_z^2}{F^4}=0$, i.e. $F(z)=\frac{1}{\alpha z+\beta}$ with $\alpha$, $\beta$ real constants.

When $\pi_{11}$, $s^2$, $\tilde{\rho}\neq 0$ the elastic properties of this solution are described in (\ref{solA13})-(\ref{solA16}), which consists of a system in the $n$-dependent variables: the three pull back material metric components $\eta_{11}$, $\eta_{22}$, $\eta_{33}$, the conformal factor $F$, and the unsheared quantities $\tilde{\mu}$, $\tilde{p}$, $\tilde{\rho}$. Notice that we can take three of the variables to be free. Working out equations (\ref{solA13})-(\ref{solA16}) gives

\begin{eqnarray}
&&\tilde{\mu}+\frac{\tilde{\rho}}{36}\left[\frac{(\eta_{11}+\eta_{22}+\eta_{33})^3-(\eta_{11}^3+\eta_{22}^3+
\eta_{33}^3)}{F^6}-24\right]=\nonumber\\
&&\frac{-2F_{zz}F+F_z^2}{F^4}\label{solA13elastic}\\
&& \tilde{p}+\frac{(\tilde{\Omega}-1)\tilde{\rho}}{36}\left[\frac{(\eta_{11}+\eta_{22}+\eta_{33})^3-(\eta_{11}^3+\eta_{22}^3+
\eta_{33}^3)}{F^6}-24\right]= \nonumber\\
&&\frac{4F_{zz}F+F_z^2}{3F^4}\label{solA14elastic}\\
&&\tilde{\rho}(\eta_{11}^2-\eta_{22}\eta_{33})(\eta_{22}+\eta_{33})=4F_{zz}F^3-8F_z^2F^2
\label{solA15elastic}\\
&&\eta_{11}^2(\eta_{22}+2\eta_{33})-\eta_{22}^2(\eta_{11}+2\eta_{33})+\eta_{33}^2(\eta_{11}-\eta_{22})=0.
\label{solA16elastic}
\end{eqnarray}

In particular, the system admits a solution when $\eta_{22}=\eta_{11}$. In this case, the material metric is conformally flat. Notice that, if $\eta_{33}=\eta_{22}=\eta_{11}$ then $\pi_{11}=0$.

 When $\eta_{22}=\eta_{11},$ equation (\ref{solA16elastic}) is trivially satisfied and the system takes the simpler form

\begin{eqnarray}
&&\tilde{\mu}+\frac{\tilde{\rho}}{36}\left[\frac{(2\eta_{11}+\eta_{33})^3-(2\eta_{11}^3+
\eta_{33}^3)}{F^6}-24\right]=\nonumber\\
&&\frac{-2F_{zz}F+F_z^2}{F^4}\label{solA13elastica}\\
&& \tilde{p}+\frac{(\tilde{\Omega}-1)\tilde{\rho}}{36}\left[\frac{(2\eta_{11}+\eta_{33})^3-(2\eta_{11}^3+
\eta_{33}^3)}{F^6}-24\right]= \nonumber\\
&&\frac{4F_{zz}F+F_z^2}{3F^4}\label{solA14elastica}\\
&&\tilde{\rho}\eta_{11}(\eta_{11}^2-\eta_{33}^2)=4F_{zz}F^3-8F_z^2F^2.
\label{solA15elastica}
\end{eqnarray}

One can solve the previous system for $\eta_{11}$, $\eta_{33}$ and $\tilde{p}$. The solution gives a relation between $\tilde{p}$, $\tilde{\mu}$, $\tilde{\Omega}$, $F(z)$ and its derivatives up to second order and both $\eta_{11}$ and $\eta_{33}$ expressed in terms of $\tilde{p}$, $\tilde{\rho}$ , $\tilde{\Omega}$, $F(z)$ and its derivatives up to second order.

 The space-times described in this subcase are inhomogeneous, rotating and possess shear but no angular velocity nor expansion. Their elastic properties are described in (\ref{solA13elastica})-(\ref{solA15elastica}).

\bigskip

{\bf Case A2*}: $\omega_\alpha=\sigma_{12}=\sigma_{22}-\sigma_{11}=0$

\bigskip

 Here the space-times are non rotating $\omega_\alpha=0$ with shearing components $\sigma_{11}=\sigma_{22}$. The only non zero acceleration term satisfies $a_3=-\dot{u}_3$. Therefore, (\ref{JE1})-(\ref{JE11}) reduce to

\begin{eqnarray}
&&e_3(\dot{u}_3)=\frac{9}{2}\sigma_{11}^2-\frac{3}{2}\dot{u}_3^2-\frac{1}{2}\Lambda-\frac{1}{2}\mu \label{A2*1}\\
&&e_3(\Omega_{3})=-\dot{u}_3\Omega_{3}\label{A2*2}\\
&&e_3(\sigma_{11})=-\dot{u}_3\sigma_{11}\label{A2*3}\\
&&\mu +\frac{3}{2}p=-\frac{27}{2}\sigma_{11}^2+\frac{1}{2}\Lambda+\frac{3}{2}\dot{u}_3^2.\label{A2*4}
\end{eqnarray}

 Introducing the simple equation of state (\ref{equationstate1})-(\ref{equationstate3}) in the previous equations we obtain the following ODE system in the variables $\sigma_{11}$, $\Omega_{3}$ and $\dot{u}_3$

\begin{eqnarray}
&&e_3(\dot{u}_3)=\frac{9}{2}\sigma_{11}^2-\frac{3}{2}\dot{u}_3^2-\frac{1}{2}\Lambda-\frac{1}{2}\tilde{\mu}-\frac{1}{2}\tilde{\rho}s^2\label{A2*1es}\\
&&e_3(\Omega_{3})=-\dot{u}_3\Omega_{3}\label{A2*2es}\\
&&e_3(\sigma_{11})=-\dot{u}_3\sigma_{11}\label{A2*3es}
\end{eqnarray}

\noindent along with the conditions

\begin{eqnarray}
&&\tilde{\mu}+\frac{3}{2}\tilde{p}+\frac{1}{2}(3\tilde{\Omega}-1)\tilde{\rho}s^2=-\frac{27}{2}\sigma_{11}^2
+\frac{1}{2}\Lambda+\frac{3}{2}\dot{u}_3^2\label{A2*4es}\\
&&\pi_{11}=-(\tilde{\mu} +\tilde{p}+\tilde{\Omega}\tilde{\rho}s^2)=\frac{\tilde{\rho}}{6F^2}\left[({\eta^\rho}_\rho )^2\eta_{\langle 11\rangle}-\eta^{\rho\sigma}
\eta_{\rho \langle 1}\eta_{1\rangle\sigma}\right]\label{A2*5es}\\
&&\tilde{\rho}\left[({\eta^\rho}_\rho )^2\eta_{\langle 22 \rangle}-\eta^{\rho\sigma}\eta_{\rho \langle 2}\eta_{2\rangle\sigma}\right]=
\tilde{\rho}\left[({\eta^\rho}_\rho )^2\eta_{\langle11 \rangle}-\eta^{\rho\sigma}\eta_{\rho \langle 1}\eta_{1\rangle\sigma}\right].\label{A2*6es}
\end{eqnarray}

By integrating the system using the coordinates and basis indicated in (\ref{basis}) we conclude that for a non null cosmological constant the only existing solution is a perfect fluid ($\pi_{11}=0$) with equation of state $\mu =-p$, the conformal factor taking the form $F(z)=\frac{1}{\alpha z+\beta}$, where $\alpha$ and $\beta$ are real constants. In the case where $\Lambda =0$ the system has a solution with $\pi_{11}\neq 0$

\begin{eqnarray}
&&\sigma_{11}=0\label{solA21}\\
&&\dot{u}_3=\frac{F_z}{F^2}\label{solA22}\\
&&\Omega_3=\frac{C_1}{F}\label{solA23}\\
&&\tilde{\mu}+\tilde{\rho}s^2= \frac{-2F_{zz}F+F_z^2}{F^4}\label{solA24}\\
&& \tilde{p}+(\tilde{\Omega}-1)\tilde{\rho}s^2=\frac{4F_{zz}F+F_z^2}{3F^4} \label{solA25}\\
&&\tilde{\rho}\left[({\eta^\rho}_\rho )^2\eta_{\langle11\rangle}-\eta^{\rho\sigma}
\eta_{\rho \langle1}\eta_{1\rangle\sigma}\right]=\frac{4F_{zz}F-8F_z^2}{F^2}\label{solA26}\\
&&\tilde{\rho}\left[({\eta^\rho}_\rho )^2\eta_{\langle22\rangle}-\eta^{\rho\sigma}
\eta_{\rho \langle2}\eta_{2\rangle\sigma}\right]=\tilde{\rho}\left[({\eta^\rho}_\rho )^2\eta_{\langle11\rangle}-\eta^{\rho\sigma}
\eta_{\rho \langle1}\eta_{1\rangle\sigma}\right],\label{solA27}
\end{eqnarray}

\noindent where $F=F(z)$ is a real $C^2$ non null function and $C_1$ is an arbitrary real constant.

Notice that equations (\ref{solA24})-(\ref{solA27}) are the same as those in (\ref{solA13})-(\ref{solA16}), so that the elastic behaviour of the solution in this subcase is the same as that of {\bf A1*}. This subcase corresponds to a shearless, non expanding, non rotating, inhomogeneous solution with vorticity.

\bigskip

{\bf Case A3*}: $\sigma_{\alpha\beta}=\Omega_3+\omega_3=0$

\bigskip

These space-times are shearless, possess rotation and angular velocity such that the only non null components satisfy $\Omega_3=-\omega_3$ and the non zero acceleration term is $a_3=-\dot{u}_3$. Here the equations are given by

\begin{eqnarray}
&&e_3(\dot{u}_3)=-3\dot{u}_3^2-2\omega_{3}^2-\Lambda+\frac{1}{2}(\mu +3p)\label{A3*1}\\
&&e_3(\omega_{3})=-\dot{u}_3\omega_{3}\label{A3*2}\\
&&\mu +\frac{3}{2}p=\frac{3}{2}\omega_{3}^2+\frac{1}{2}\Lambda+\frac{3}{2}\dot{u}_3^2.\label{A3*3}
\end{eqnarray}

\bigskip

 Integration of this system using the coordinates and basis indicated in (\ref{basis}) shows that the only solution for which  $\pi_{11}\neq 0$ has $\omega_3=0$, being therefore a subcase of {\bf A1*} with no shear nor rotation.

\bigskip

{\bf Case B}: ${\sigma}_{11}= {\sigma}_{22}={\sigma}_{12}=\dot{u}_1=\dot{u}_2=\omega_3=\omega_2+{\sigma}_{13}=\omega_1-{\sigma}_{23}=0$

\bigskip

The condition (\ref{b1}) obtained from the Bianchi identities seems to suggest that another case exists when ${\sigma}_{11}= {\sigma}_{22}={\sigma}_{12}=\dot{u}_1=\dot{u}_2=\omega_3=\omega_2+{\sigma}_{13}=\omega_1-{\sigma}_{23}=0$. However when writing the full system (Bianchi, Jacobi and Einstein equations) of equations in the coordinates and basis indicated in (\ref{basis}), we must consider that in this basis $\mu+p+\pi_{11}=0$ as we show in subsection \ref{special}.
We then conclude that this case represents a subcase of {\bf A*}.

\bigskip

\section{Summary}
\label{sec:summary}

We have studied conformally flat spacetimes with a conformal factor depending on only one spatial coordinate, considering a stress energy tensor
having diagonal anisotropic pressure tensor with only one independent component. The Bianchi identities and commutators in the 1+3 formalism give an ODE in the pressure and energy
density along with conditions for the 1+3 kinematical quantities. The Jacobi identities and Einstein equations for a very simple
case are presented. The commutators applied to the directives of the quantities involved give no extra information. These equations (Bianchi and Jacobi identities,
Einstein equations and commutators) form the complete system that characterises this particular class of solutions. For non zero cosmological constant we obtained only a perfect fluid solution for a particular conformal factor. Solutions to this system
for which the anisotropic
pressure tensor is non zero are obtained. In fact such solutions only exist for null cosmological constant. The elastic behavior of these solutions can
be characterised by a simple equation of state where the energy per particle density depends only on the shear scalar which is an invariant
of the material metric. Unfortunately, all solutions obtained here have null expansion. We suspect that among the wider class of spacetimes described in case {\bf A} it might be possible to find solutions with expansion. However, dealing with the 1+3 complete system of equations, in that case, is a much harder problem.

\bigskip

{\bf Acknowledgements}

\bigskip

This research was partially supported by the Research Centre of Mathematics of the University of Minho through the FEDER Funds "Programa Operacional Factores de Competitividade COMPETE", and by the Portuguese Funds through FCT - "Funda\c{c}\~{a}o para a Ci\^{e}ncia e Tecnologia" within the Project Est-C/MAT/UI0013/2011. IB also thanks CMAT for support through the FCT Project Est-OE/MAT/UI0013/2014.


\bigskip




\bigskip

\begin{thebibliography}{99}

\bibitem{weber} J. Weber. \newblock {\it Phys. Rev.}, \textbf{117}, 306 (1960)

\bibitem{carter} B. Carter and H. Quintana. \newblock {\it Proc. R. Soc. A}, \textbf{331}, 57 (1972)

\bibitem{souriau} J.M. Souriau. \newblock {\it G\'{e}om\'{e}trie et Relativit\'{e}}, (Paris: Herman) (1965)

\bibitem{maugin1} G.A. Maugin.\newblock {\it Ann. Inst. H Poincar\'{e} A}, \textbf{15}, 275 (1971)

\bibitem{maugin2} G.A. Maugin.\newblock {\it J. Math. Phys.}, \textbf{19}, 1212 (1978)

\bibitem{magli1} J. Kijowski and G. Magli. \newblock {\it Geom. Phys.}, \textbf{9}, 207 (1992)

\bibitem{magli2} J. Kijowski and G. Magli. \newblock {\it Class. Quantum Grav.}, \textbf{15}, 3891 (1998)

\bibitem{christo} D. Christodoulou. \newblock {\it Ann. Inst. H Poincar\'{e} A}, \textbf{69}, 335 (1998)

\bibitem{beig} R. Beig and B. G. Schmidt. \newblock {\it Class. Quantum Grav.}, \textbf{20}, 889 (2003)

\bibitem{karl1} M. Karlovini and L. Samuelsson.\newblock {\it Class. Quantum Grav.}, \textbf{20}, 3613 (2003)

\bibitem{karl2} M. Karlovini and L. Samuelsson.\newblock {\it Class. Quantum Grav.}, \textbf{21}, 1559 (2004)

\bibitem{karl3} M. Karlovini and L. Samuelsson.\newblock {\it Class. Quantum Grav.}, \textbf{21}, 4531 (2004)

\bibitem{karl4} M. Karlovini and L. Samuelsson.\newblock {\it Class. Quantum Grav.}, \textbf{24}, 3171 (2007)

\bibitem{park} J. Park.\newblock {\it Gen. Rel. Grav.}, \textbf{32}, 235 (2000)

\bibitem{irene} I. Brito, J. Carot and E. G. L.R. Vaz. \newblock {\it Gen. Rel. Grav.}, \textbf{42}, 2357 (2010)

\bibitem{calo} S. Calogero and J. M. Heinzle.\newblock {\it Class. Quantum Grav.}, \textbf{24}, 5173 (2007)

\bibitem{uggla} H.V. Elst and C. Uggla.\newblock {\it Class. Quantum Grav.}, \textbf{14}, 2673 (1997)

\bibitem{ehlers} J. Ehlers.\newblock {\it Gen. Rel. Grav.}, \textbf{25}, 1225 (1993)

\bibitem{wain} J. Wainwright and G. F. R. Ellis. \newblock {\it Dynamical Systems in Cosmology}, (Cambridge: Cambridge University Press) (1997)

\bibitem{misner} C. W. Misner, K. S. Thorne and J. A. Wheeler. \newblock {\it Gravitation}, (New York: Freeman) (1973)

\bibitem{ellis} G. F. R. Ellis. \newblock {\it Proc. ICGC95 Conf. at IUCCA, Pune, ed S. Dhurandhar and T. Padmanahan}, (Dordrecht: Kluxer) (1997)

\bibitem{np} E. Newmann and R. Penrose. \newblock {\it J. Math. Phys.}, \textbf{3}, 566 (1962)

\bibitem{wald} R. M. Wald. \newblock {\it General Relativity}, (Chicago: University of Chicago Press) (1984)

\bibitem{felice} F. de Felice and C. J. S. Clarke. \newblock {\it Relativity on curved manifolds}, (Cambridge: Cambridge University Press) (1990)

\bibitem{edgar} S. B. Edgar. \newblock {\it Gen. Rel. Grav.}, \textbf{12}, 347 (1980)

\bibitem{pirani} F. A. Pirani.\newblock {\it Acta Phys. Polon.}, \textbf{15}, 389 (1956)

\bibitem{ellis1} G. F. R. Ellis. \newblock {\it J. Math. Phys.}, \textbf{8}, 1171 (1967)

\bibitem{mac} M. A. H. MacCallum. \newblock {\it Cosmological Models from a Geometric Point of View}, (New York: Gordon and Breach) (1973)
    
\bibitem{ellismac}G. F. R. Ellis and M. A. H. MacCallum. \newblock {\it Commun. Math. Phys.}, \textbf{12}, 108 (1969)


\end{thebibliography}
\end{document}